# Mechanism on Failure of Cooled Infrared Detector Illuminated by CW laser

Zi-Yi Zhang, Chen-Wu Wu

Institute of Mechanics, Chinese Academy of Sciences, Beijing, 100190, China

University of Chinese Academy of Sciences, Beijing, 100049, China

Corresponding address: No. 15 Beisihuanxi Road, Haidian District, Beijing 100190, China. Email: chenwuwu@imech.ac.cn & c.w.wu@outlook.com

**Abstract**:

Cooled infrared detectors with high sensitivity and high performance are widely applied in many fields. However, environmental disturbances such as intense light may cause a decline in their performance and even lead to permanent damage. In this study, the multi-physical behaviors and the performance degradation processes of typical cooled infrared detectors under the illumination of high-power density lasers are investigated to evaluate their survivability under extreme conditions. The experimental results reveal the electrical and thermomechanical responses of the detectors under the illumination of continuous-wave lasers, and the main thermomechanical failure modes that have not been revealed before are discovered. The theoretical results obtained through a newly established three-dimensional multi-physical numerical model of the detectors under laser illumination clarify the thermomechanical mechanism of the failure of the main components of the detectors. Moreover, the dependence of the thermomechanical results on relevant parameters is discussed, providing a scientific basis for the development and optimization of related systems.

## 1. Introduction:

Cooled infrared detectors, a sort of photoelectrical devices which convert incident infrared radiation into electrical signals, are widely used in many fields such as biomedical [1], environmental [2], manufacture [3], aerospace engineering [4] and so on. Infrared detectors are generally classified into single element detectors, linear array and focal plane arrays according to the distribution of photoelectrical chips. It is a promising trend to develop low-cost, miniature, and excellent performance detectors in the future [5]. Infrared detectors could also be divided into near-infrared, mid-infrared and far-infrared detectors based on the optimum operation wave band. Many kinds of chip materials are designed to accommodate various service environment and detection targets. For example, lead-salt semiconductors are commonly used in the near-infrared (NIR) band, while Mercury Cadmium Telluride (HgCdTe), Indium Antimonide (InSb), and type-II superlattices are more frequently used in the mid-infrared and far-infrared (FIR) bands [6]. Particularly, Indium Antimonide (InSb) and mercury cadmium telluride (HgCdTe), applied in high-speed aircrafts, exhibit high quantum efficiency behavior operating at low temperature. However, detectors made of these materials demand a more stable temperature condition as they are more sensitive to temperature fluctuations. It is highly possible that their performance would be affected greatly when exposed to the illumination of strong light and high temperature [7], which also provides the thermodynamic basis for the electro-optic countermeasure technology. To create more adaptable and durable detectors [8-10] for future applications, researchers around the world have been dedicated to development of new structures [11] and chip materials [12, 13] which contribute to excellent operational performance under severe conditions. Therefore, studies on damage mechanism of high-performance infrared detectors under laser illumination have a significant meaning for the development of the electro-optic countermeasure technology. By recognizing the damage evolution process of the detector's destruction and further seeking out the critical parameters or damage threshold in system design, not only could the detector's environmental adaptability and survivability be enhanced, but also higher efficiency could be achieved in interfering and destroying photoelectric equipment in electro-optic countermeasure tasks.

Different parts of the cooled infrared detector would behave differently when illuminated by laser whose photonic energy exceeds the band gap of the chip material. The laser illumination on the chip would intrigue electronic, thermal and mechanical responses simultaneously, while leading to thermal and mechanical response only on the optical window. Based on previous research according to our knowledge, most of the previous studies only focus on the laser-induced damage mechanism of detector chip. Furthermore, it is commonly reported that there could be fundamentally two categories of laser-induced damage to the detector. One is known as soft fault which is manifested as nonlinear electrical response like saturation [14], supersaturation [15], memory effect [16], zero voltage output [17], etc. The other is mechanical damage of the detector such as the melting and fracturing of the detector chip, or the breaking of leading wire [18-21].

For infrared detectors with high sensitivity and precision, the infrared optical window not only serves the function of allowing target radiation to pass through but also plays a critical role in filtering stray light. Furthermore, optical windows are indispensable in high-power lasers [22,23], making it essential to conduct in-depth research into the laser damage mechanisms of infrared components and to enhance the damage threshold of laser materials through structural design [24]. Its resistance to damage directly determines the performance and service life of the entire device. This is crucial for optimizing the performance of optoelectronic devices such as infrared detectors and high-power lasers, etc. With relatively high compressive strength and low plasticity, the optical window is prone to damage under high-power-density laser illumination. The laser-induced damage in the optical window is predominantly attributed to two factors: significant temperature elevation [25] and localized stress concentration [26]. These phenomena are typically induced by local light absorption peaks resulting from randomly distributed impurities and defects.

Optical window typically exhibits a multilayer heterogeneous structure, composed of a relatively thick central bulk material and a thin surface optical films. As a result, their damage modes under laser illumination can be categorized into bulk damage and surface damage [24]. Laser-induced damage to optical window primarily manifests as material melting [27], deterioration, or fracture caused by high temperatures and stress. Factors such as material bandgap [28], processing techniques [29], surface defects [25], contamination, and laser parameters [30] significantly influence the laser damage threshold. Some researchers have also found that the deposition temperature significantly influences the surface morphology and the magnitude of residual stress in thin films, thereby further affecting the performance of optical window [31]. Previous studies have established thermo-mechanical coupled finite element models to validate the thermo-mechanical damage process dominated by thermal stress under surface contamination conditions [32, 33].

The semiconductor chip of infrared detector influenced by laser illumination undergoes processes at the micro level such as carrier excitation, recombination, transportation. In certain cases, the performance of semiconductor material would be enhanced greatly by the application of laser processing during the manufacturing of the detector. For instance, laser annealing is commonly used to remove impurities and defects in the semiconductor-material [34]. However, in many circumstances, intensified laser illumination would result in the non-reversible damage and even permanent failure of the infrared detector chips. The degree and characteristics of damage caused by laser illumination depends largely on not only the physical and chemical properties of semiconductor material but also laser parameters such as wavelength, pulse width, pulse number and power density [35], etc.

The dominant mechanisms of damage to the detector chip by in-band laser, which means that the laser wavelength fall into the optimum operation wave band of the detectors, are thought to be photoelectric and thermal effects. For out-of-band lasers, photoelectric and thermal effects occur simultaneously only when laser wavelength is shorter than the optimum operation wave band. When laser wavelength is longer than the upper limit of the band, only thermal effects would arise in the detector chip [36]. The damage thresholds of the detector chip under different optical parameters of laser beams have been systematically analyzed and summarized for specific conditions in previous studies. It is usually believed that the detector chip is easier to be damaged utilizing the pulsed-laser than the continuous wave laser if the total incident energy is roughly equal. What's more, it is reported that the damage threshold would decrease with the increasing repetition frequency of pulsed-laser. In addition, Zhang, et al. declared that the laser damage thresholds of Indium Antimonide (InSb) infrared detectors are correlated with the laser's pulse width [37]. Christopher Burgess, et al. further elucidate the different mechanism for laser damage thresholds of typical infrared detector under different illumination duration by adopting numerical modeling method [38].

Except for the laser parameters, the material compositions and performance parameters of the semiconductor chip also have a great influence on failure modes of the detector chip and it's laser damage thresholds. Detector chip made of different material suffers from laser illumination in different ways. For instance, illumination on chips made of Cadmium Zinc Telluride (CZT) and Mercury Cadmium Telluride (CdTe) usually lead to the evolution behavior of mercury at high temperatures. However, cracks are found to arise in the Cadmium zinc telluride chip (CZT) rather than Mercury Cadmium Telluride (HgCdTe). Even made up of the same materials, the laser damage threshold of photovoltaic detectors is usually believed to be higher than photoconductive ones [39].

All of the aforementioned studies might have preconceived that the failure process of the infrared detectors under laser illumination is overwhelmingly dominated by the destruction of the chip with high temperature elevation. Based on this assumption, numerous experiments and numerical models have been carried out to explore the detector chip's damaging process, such as melting, vaporization, or the thermal decomposition with temperature rising [40]. As for the cooled infrared detectors, failure of the packaging and interconnecting structure due to thermal stress is one of the most important reasons for its laser-induced damage. The multi-layered chip undergoing thermal cycle repeatedly is prone to become a high stress region owing to the varied temperature gradient or mismatch in thermal expansion between different layers, leading to the failure of the interlayer connection [41]. Because of limited experimental conditions and high cost for the destructive testing of the cooled infrared detectors, numerical modeling method has been widely used to analyze the stress state of detector chip illuminated by laser [42, 43]. By means of the numerical computations, different technical suggestions are proposed to reduce thermal stress during the refrigeration process [44, 45].

In most previous research on multi-physical behaviors of the cooled infrared detectors under laser illumination, the mathematical-physical model of the detector chip is established in a one-dimensional way, in which the temperature variation of any other component surrounding detector chip is not taken into consideration. Most studies on thermal stress in infrared windows primarily focus on analyzing the window itself, without considering the window as an integral part of the overall infrared detector system [46]. However, it's known to all that the validity of vacuum insulation layer in a liquid nitrogen cooled infrared detector highly depends on the integrity of the optical window. Such kind of cooled infrared detector usually consists of the chip and a Dewar composed of optical window, cold finger, inner cylinder and outer cylinder. The Dewar not only provides optical, electrical, and mechanical interfaces for the detector chip, but also serves to maintain a high vacuum and low temperature state to ensure the normal operation of the detector chip [41]. Obviously, the structural integrity and effectiveness of the Dewar's vacuum layer is an indispensable prerequisite for the work of the detector. Thus, it is obviously biased to judge the reliability of the detector by focusing only on thermomechanical damage of the chip without enough consideration on the optical window.

This study aims to analyze high-performance infrared systems as an integrated whole and elucidate their failure and damage mechanisms under extreme environments based on thermo-mechanical theory, thereby providing a foundation for the optimization of high-performance infrared systems. In the present work, a series of experiments on liquid nitrogen cooled infrared detectors laser illuminated by laser with different power density are designed and carried out. The open-circuit voltage of the detector and temperature at its outer surface is monitored in the entire process of the experiment. When the laser power density has reached to the level to cause permanent anomalies in the voltage and temperature responses, laser is turned off and the morphologies of the detector chip and optical window are observed and analyzed under the microscope. The failure mode and its critical parameters interval of the cooled infrared detector is revealed by the experimental results. After that, the thermomechanical responses of the detector system to laser illumination are analyzed by adopting the finite unit method in the numerical modeling. The temperature, deformation and stress fields of the whole structure of the detector illuminated by laser are solved the appropriate thermomechanical boundary conditions to simulate the in-situ test state. The numerical calculations results elucidated the failing process and its parameter dependency with the cooled infrared detector illuminated by continuous wave laser. Accordingly, a crucial perspective on failure mechanism of liquid nitrogen cooled infrared detectors subjected to continuous wave laser illumination is established on the comprehensive understanding of the experimental and numerical results.

## 2. Experimental description and results:

The experimental set-up established in present work is shown in Fig.1, the cooled InSb photovoltaic detector is illuminated by a continuous wave laser of wavelength 808nm, and the signal output end of the detector is connected to the oscilloscope through a Bayonet Nut Connector (BNC) so that open-circuit voltage change of the detector can be monitored in real time. The detector is mainly composed of a photoelectric chip unit and a liquid nitrogen Dewar which maintains the low-temperature environment for the chip unit. The photoelectric chip is designed to work in the optimum operation wave band of 2~5um, whose performance parameters are shown in Table 1.

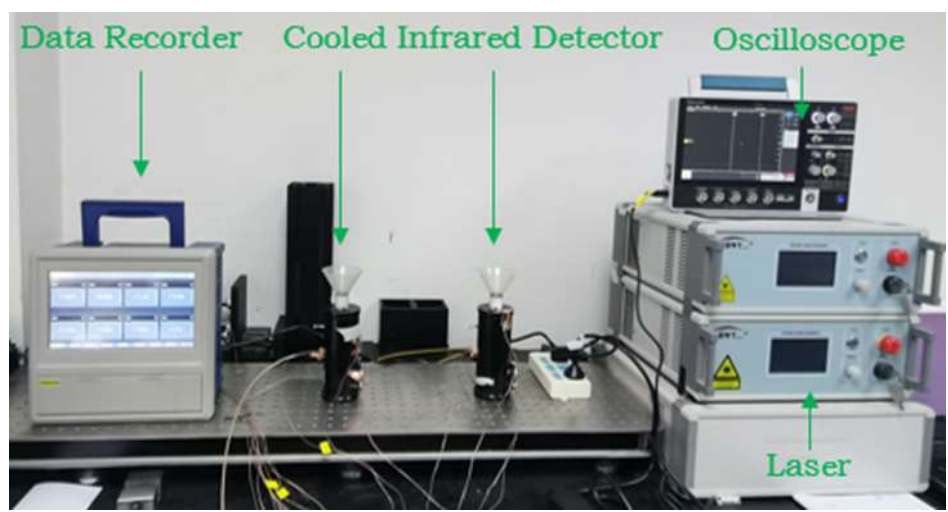

Fig. 1 Diagram of the experimental set-up made up of Laser, Oscilloscope, Data recorder and the Cooled Infrared Detector under test

Table 1 performance parameters of the InSb detector

| wave band μm | pixel size mm² | Peak responsivity $(\frac{A}{W})$ | Time constant(μs) | Peak detectivity $(cm \cdot Hz^{\frac{1}{2}} \cdot W^{-1})$ |
|---|---|---|---|---|
| 2~5 | 1 × 1 | ≥ 2.0 | < 1 | $8 \times 10^{10}$ |

In the experiment, the laser beam is vertically incident on the detector window with the spot diameter of 2mm to cover the whole surface of the photosensitive unit. The duration of laser illumination is about 3s, and enough time interval is reserved to restore the chip temperature to the pre-illumination state before next test with the gradually increased laser power density.

The tests reveal that, when the laser power exceeds 5 watts, the discrepancy between the set values and the actual values is less than 10%. This is in consistency with the systematic comparative analysis between the set values and the actual values of laser power of the same equipment in prior studies in the same laboratory [47].

By adjusting the sampling rate of the oscilloscope, the time span of data recording is set to cover the whole process of laser illumination, ensuring that the changes in output voltage of the detector throughout the whole process (before, during and after illumination) are completely recorded. During the test, the laser power is gradually increased to change the output open-circuit voltage that is monitored in real time. The open-circuit voltage of the detector is sustained to be around 0V at room temperature. Before laser illumination begins, liquid nitrogen is injected into the inner cylinder of the Dewar through a funnel to decrease the chip temperature to its operating temperature range. The open-circuit voltage of the detector remains about 80 mV when the temperature of the detector chip is within the operating temperature range. According to time sequence during the experimental process, every single test of laser illumination can be divided into three stages: 1st stage: before illumination (*ca*. 0~2s), 2nd stage: during illumination (*ca*. 2~5s), and 3rd stage: after illumination (*ca*. 5~7s). The curve in Fig. 2(a) depicts the open-circuit voltage variation process throughout an entire test. In the 1st stage, it is shown that the open-circuit voltage stabilizes at 80 mV before illumination (*ca*. 0-2s). In the 2nd stage, the open-circuit voltage increases sharply to 130mV at the instant the laser is activated. It then reduces gradually to a relatively steady value after one to two seconds of laser illumination. At the beginning of the 3rd stage, the open-circuit voltage decreases abruptly at the instant the laser is turned off. Then it increases gradually before it returns to 80mV.

As shown in Fig. 2(b), it could be clearly observed that the magnitude of abrupt change at the start-up and shut-down instant of laser basically remains equivalent. Consequently, the maximum open-circuit voltage of the detector during illumination, which arises from a value of about 80 mV at the start-up instant, always falls near 130mV with the laser power ranging from 0.5W to 24W according to the data collected. However, the time required for the open-circuit voltage to decrease to the steady value extends from 2s to 4s with the increasing laser power. At the same time, the steady value in the 2$^{nd}$ stage decreases, leading to the minimum open-circuit voltage decreasing from about 80 mV to 20mV. But under the effect of cooling system after illumination (*ca*. 6-7s) in the 3$^{rd}$ stage, the open circuit voltage finally returns to the same value(80mV) as in the 1$^{st}$ stage.

Experimental results demonstrate that the steady-state open-circuit voltage exhibits a decreasing trend with increasing laser power within range from 0.5 to 24W, which leads to the conclusion that the detector operates in a state of output over-saturation beyond its optimal range. The open-circuit voltage, which represents the maximum output voltage that a photovoltaic detector can provide to an external circuit [48], mainly comprises two components, the voltage induced by photon and the voltage induced by temperature elevation [49]. Based on the evidence above, it could be deduced that the decrease of open-circuit voltage in the 2$^{nd}$ stage is primarily due to the rise in chip temperature.

When the detector chip reaches a thermodynamic equilibrium state around its operating temperature within the liquid nitrogen Dewar, the open circuit voltage stabilizes at 80mV. Once the thermal boundary conditions are altered by the influence of high power density laser, the chip temperature as well as its open-circuit voltage immediately changes. The shape of the voltage curves is close to a square wave when the laser power is lower than 1W as the temperature variation is not obvious under these conditions. When the laser power increases, the temperature variation becomes larger. Therefore, it's more difficult for the system to stabilize at initial state, and the time spent on reaching a new equilibrium state would be longer.

The sequential process of saltation and stabilization in open-circuit voltage curve, which is triggered by the start-up and shut-down of laser, is described as the "double-ear" [15, 50, 51] phenomenon. This process is a typical over-saturation phenomenon of photovoltaic (PV) type detectors under continuous laser illumination. The abrupt variations observed in the open-circuit voltage at the starting moment of the 1$^{st}$ and 3$^{rd}$ stages are likely attributable to the rapid fluctuations in carrier concentration induced by the start-up and shut-down of the laser. And the gradual decrease of the open-circuit voltage before stabilization during the 2$^{nd}$ stage is due to the cumulative heating of the detector chip caused by prolonged laser illumination, as previous research has demonstrated that the open-circuit voltage gradually decreases as the operating temperature increases [49]. When the laser power is maintained at a relatively low level, the temperature increase induced by laser illumination can be rapidly dissipated by the liquid nitrogen Dewar. When the laser power gradually increases, the temperature rise of the chip gets more pronounced. It takes longer for the carriers to reach thermal equilibrium again, so the steady value of open-circuit voltage in the 2$^{nd}$ stage becomes smaller [52, 53].

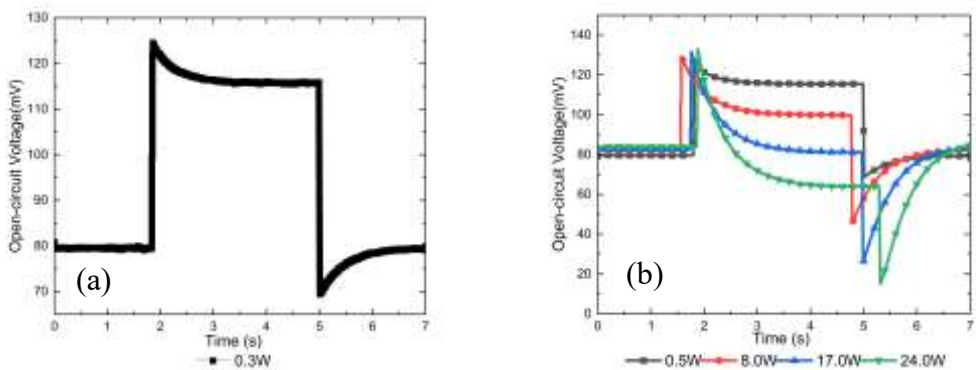

Fig. 2 (a) Typical characteristic of time history of the open-circuit voltage of the detector during illumination and (b) Time history of the open-circuit voltage of the detector when being illuminated by laser of power 0.5W, 8.0W, 17.0W and 24.0W

When the laser power reaches 27 W, the detector undergoes permanent failure, accompanied by an audible cracking sound, after 5 seconds of illumination. Visual inspection revealed distinct cracks on the optical window's exterior surface, with the surface films in the central laser spot region exhibiting significant carbonization. Following a period of cooling, the temperature of the detector's outer cylinder decreases below 0°C, leading to progressive frost formation that eventually results in complete snow coverage across the entire external surface of the device, which is shown in Fig. 3(a) with the left one for global diagram and right one for partial enlarged view.

In the meantime, the open-circuit voltage of the detector decreases quickly to 0 mV. Although subsequent tests have indicated that an open-circuit voltage could be intrigued again with liquid nitrogen injected into the inner cylinder of the Dewar, but it is always below 80 mV under similar ambient radiation, which means that the detector chip is not working properly at its operation temperature. Moreover, this lower open-circuit voltage soon returns to 0 as the chip temperature rises rapidly again, which verifies that breakdown of the vacuum layer within the Dewar has led to fatal degradation of its cooling capacity. Based on the facts mentioned above, it can be concluded that the detector has failed permanently. Finally, the optical window and the detector chip is taken out from the disassembled detector for observation under optical microscopy and scanning electron microscopy (SEM).

The surface films in the illuminated zone is completely burned on the exterior surface, which is in the form of black paste, as shown in Fig. 3(b). Furthermore, the material in adjacent area surrounding the illuminated zone has also been melted and deteriorated to a certain extent as shown in Fig. 3(c). It can be deduced that the temperature on the exterior surface rises more substantially than that on the interior surface. This distinction might be resulted from the impurities and pollutants [25] attached to the exterior surface as shown in Fig. 3(d), which would form a lot of localized absorption peaks of the laser near the contaminated area. According to the research of C. R. Haas et al [54], temperature rises rapidly near the pollutants and the process of ablation will be speeded up with the chemical reaction between the coating films material, water and oxygen in the air at high temperatures. Ultimately, there are craters caused by the ablation of material on the outer surface of the optical window, as shown in Fig. 3(e).

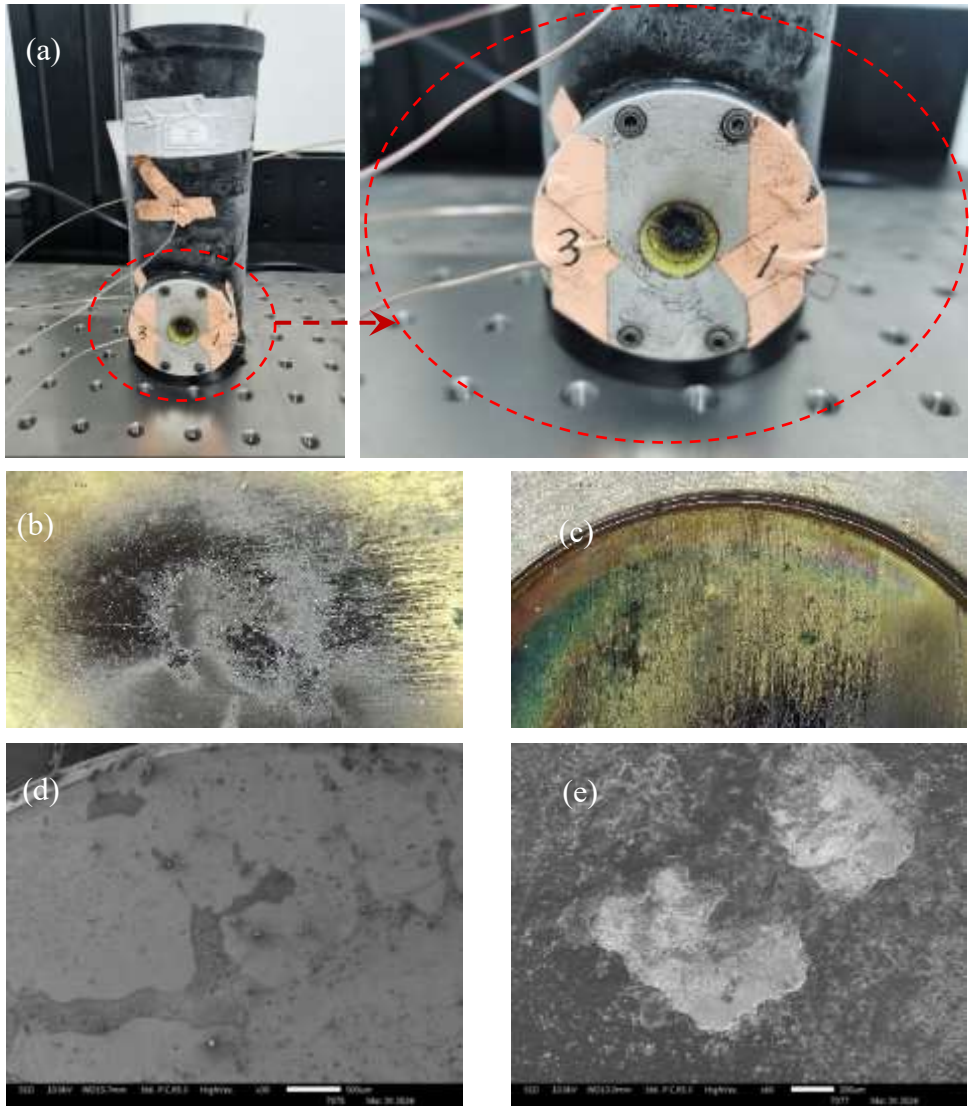

Fig. 3 (a) Frost on the outer surface of the detector after the vacuum break down; the typical damage morphology upon exterior surface of the optical window: (b) color darkening in the central of the optical window directly illuminated area, (c) melting of the surface films of the optical window, (d) pollutants on the outer surface and (e) craters on the outer surface of the optical window

The optical window of the cooled infrared detector is a kind of multi-layered material with thin films of 2.25um thickness coated on both exterior and interior surface of the Zinc Selenide substrate (ZnSe). Specifically, the surface films on each side is composed of two layers of anti-reflective materials, including the colorless transparent crystal Ytterbium fluoride(YbF3) and the yellow transparent crystal Zinc

Selenide (ZnSe) layer. The melting point of Ytterbium fluoride and Zinc Selenide is 1157℃ and 1525℃ respectively. From the perspective of appearance, such optical window in study is a piece of smooth and yellowish transparent glass in the shape of a flat-disc before destruction, as shown in Fig. A1. After being permanently damaged by the laser illumination, the central region on the exterior surface appears to turn black and three main cracks intersect near the center of illuminated region, as shown in Fig. 4(a). It is revealed by Figs. 4(b)~4(d) that the angle of any crack to the another among these three cracks is approximately 120°. However, the interior surface of the optical window appears to maintain smooth and homogeneous with no obvious ablation phenomenon observed, while a piece of glass fragment detaches from the whole glass at the intersection point of the cracks.

Physically speaking, the aforementioned damage patterns of the optical window should largely attribute to the rapid temperature elevation and high thermal stress developed by laser illumination. These include the melting/ ablation as well as the fracture of the films on the exterior surface of the optical window, the interface delamination between the surface films and the glass substrate, and cracking of the zinc selenide (ZnSe) glass substrate. The melting/ ablation of surface films will lead to the change of light absorptivity of the optical window severely, the cracking of the glass substrate and interfaces between the substrate and films has significant influence on the transitivity and absorptivity of light. As a consequence, the optical performance of the detector will be altered apparently due to a dramatic change in the optical property of the detector's optical window.

The detector chip is also observed under the optical microscope and the scanning electron microscope (SEM) for comparison. As shown in Fig. 4(e) with the left one for global view and right one for partial enlargement, the surface of the detector chip is homogeneous and smooth with neither cracks nor surface color change resulted from laser illumination on it. Except for the missing corner broken from the detector chip due to the processing of disassembly, there is no obvious damage caused by laser illumination on it. All these evidences prove that when the detector permanently fail after test, mechanical damages occur on the optical window only, while the detector chip stays intact before the disassembly. It could be concluded that the permanent failure of the cooled InSb infrared detector is owing to the breakdown of the vacuum layer sustained by the integrity of the whole structure and destructed by the cracking of the optical window by the action of both high temperature elevation and thermal stress.

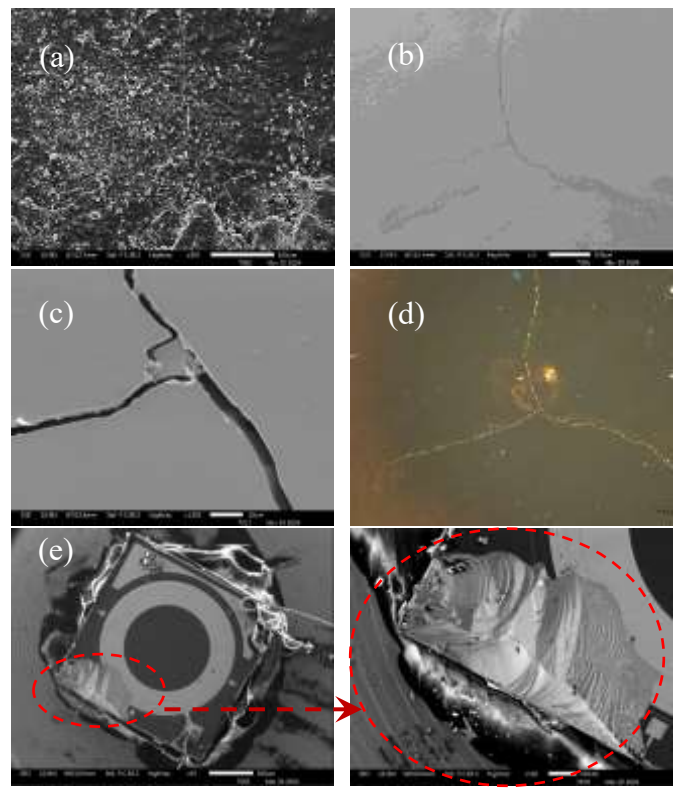

Figure. 4 The cracking morphology of the optical window: (a) the exterior surface, (b) the interior surface, (c) the glass fragment around the intersection point of the cracks by SEM and (d) the interior surface by OM; and (e) the surface morphology of the corner breakage of the detector chip caused by disassembly by SEM for comparison

## 3. Numerical model and Analysis

The photovoltaic effect and the photothermal effect take place simultaneously inside the semiconductor chip when the cooled InSb infrared detector is illuminated by a continuous wave laser with the wavelength less than the upper limit of the operation band. The photoelectric effect process could be briefly described as following: carriers in the p-n junction of the semiconductor are excited by the laser photons whose energy is higher than the band gap of InSb to produce electrons and holes, which drift to the N and P regions respectively under the influence of the built-in electrical field. By this way, the p-n junction thus acquires an additional potential difference (the photovoltage), which is shown in the increase of the open-circuit voltage of the detector measured by the oscilloscope. During the action of the photothermal effect in the detector, the thermal equilibrium is broken when it is illuminated by the laser. Through a complicated process including scattering and recombination of the excited carriers, the energy of the laser photons absorbed by the detector chip is ultimately dissipated to generate temperature elevation of the detector materials.

The laser damage of cooled infrared detector is usually caused by the high temperature elevation of the optical materials due to the photothermal effect with laser illumination. The drastic temperature elevation could make these optical materials undergo phase change when it reaches the critical temperature or experiences a series of chemical processes at high temperature, which could cause a dramatical change in the optical parameters of the material. A three-dimensional numerical model is established in present work to elucidate the temporal and spatial distribution of temperature, deformation and stress fields in the cooled infrared detector under the thermos-mechanical effects of the laser illumination.

During the course of experiment, the laser is turned on after the temperature and stress of the detector structure reach a steady state with the liquid nitrogen filled in the inner cylinder of the Dewar. Accordingly, the numerical simulation is divided into two computational steps in which the first step is aimed at solving the steady state of the Dewar's cooling process and the second step to simulate the process during and after illumination. The results of the first step provide the stationary spatial distribution of temperature, deformation and stress filed when the detector reaches the thermodynamic equilibrium state after the inner cylinder of the Dewar is filled with liquid nitrogen. According to the open-circuit voltage curves recorded during the experiment, the laser illumination duration was merely 3 seconds (*ca*. 2-5s) when the detector remained intact, whereas it's extended to 5 seconds upon detector failure. Consequently, a 5-second illumination period was adopted for the numerical simulation. The time duration of the transient solution step is set to a total length of 8s, with 0-5s for the stage of laser illumination and 5-8s for the restoring stage after illumination.

As shown in Figs. 5(a), (b) and (c), the cooled InSb infrared detector is mainly composed of four parts: the outer and the inner cylinder, connecting components between the outer cylinder and the inner cylinder, the cold finger and the semiconductor chip. The materials of the four parts aforementioned are stainless steel, aluminum alloy, Bakelite, brass, and indium antimonide in sequence. This could also be partially found in Fig. A2 with the photograph of the detector with removal of the optical window.

The discrete mesh model is shown in Figs. 5 (d), (e) and (f), in which the contact interfaces between the adjacent components are assumed to be smooth and continuous ideally for both thermal and mechanical analysis. This approach may partially obscure the non-uniform geometric characteristics of actual interfaces, potentially leading to three critical analytical deviations: 1) artificial enhancement of effective thermal conductivity, 2) underestimation of thermal gradient magnitude, and 3) inadequate characterization of stress concentration phenomena. Although, the numerical model in present work has taken into account of the nonlinear relationship between the thermo-physical parameters of the materials and their thermodynamic temperature. And the temperature dependences of the parameters including density, heat capacity, thermal conductivity, thermal expansion coefficient, elastic modulus and Poisson's ratio are shown in Fig. A3.

Specifically, it is known that the thermal conductivity of brass is much higher than that of the other materials and its variation with temperature is relatively small according to the references [55]; similarly, the variation of zinc selenide's Poisson's ratio with temperature is relatively negligible [55]. Therefore, change curves of these two parameters are not cited in Fig. A3.

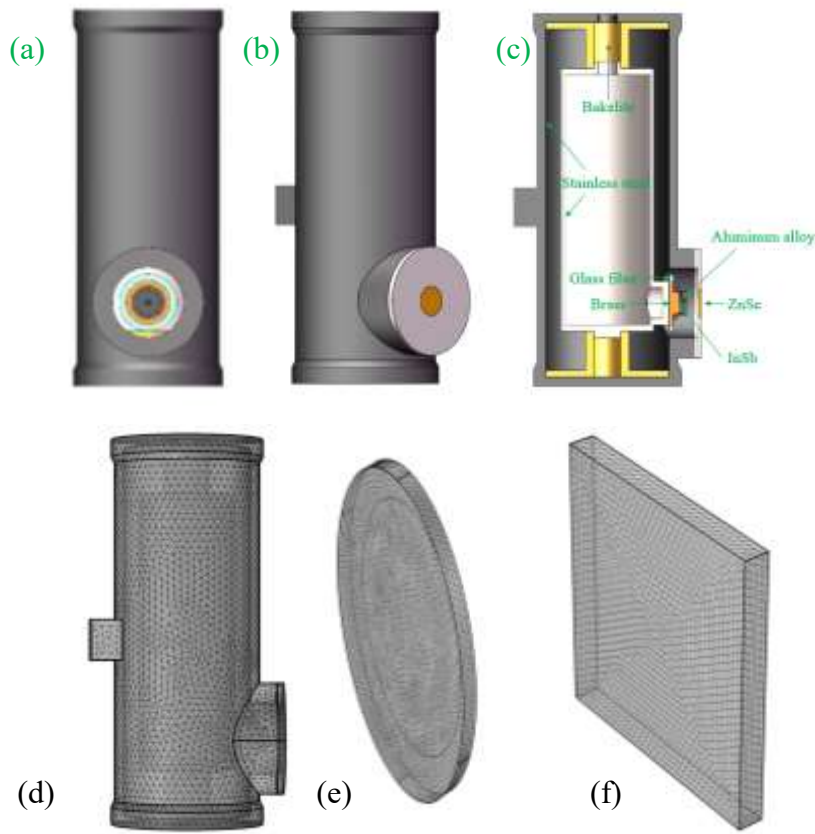

Fig. 5 The diagram of the geometrical structure: (a) inner parts behind the optical window, (b) outer surface and (c) cutaway view; The diagram of the numerical discrete mesh: (d) whole structure, (e) window glass and (f) sensor chip.

The mathematic-physical model on the variations of the fields of temperature, elastic deformation and thermal stress fields could be found in the books [57, 58] and the governing equations with specific thermo-mechanical boundary conditions are described briefly by the Equations (Eq.1) ~ (Eq.8) in the Appendix with the symbols defined in Table A1 for the Nomenclatures.

With the present numerical model, the temperature and stress results of different components are analyzed under illumination of laser with the power 27 W. Moreover, dependence of the detector's thermomechanical responses on the laser power is investigated by comparing the temperature and stress results when the laser power varies from 1W to 27 W, for which the corresponding laser power density is from $3.18 \times 10^5 W/m^2$ to $8.59 \times 10^6 W/m^2$.

The numerical results reveal that the optical window experiences the highest temperature elevation among all components of the detector after being illuminated for 5s, which is followed by the chip, the cold finger and other components connected to them secondarily, as shown in Fig. 6(a). The optical window (zinc selenide) is heated rapidly with being directly exposed to laser illumination. In the computational domain of the optical window, the temperature rises more rapidly within the regions covered by laser spot than the regions outside it till the shut-off moment of illumination. Moreover, the highest temperature elevation appears at the center of its exterior surface. While it is found that the temperature elevation in the region near interfaces between the optical window (zinc selenide) and the window frame (stainless steel) is rather small, as shown in Fig. 6(b).

Due to the considerable heat capacity of the liquid nitrogen in the inner cylinder of the Dewar, the temperature of components which are directly connected to the inner cylinder (e.g., the chip substrate) has been maintained at a level slightly above 77 K, as shown in Fig. 6(c). Computational results show that when the detector is illuminated by laser with power of 27 W, the center point of the exterior surface on the optical window reaches the temperature of 1503 K at the instant t=5s, which is close to the melting point of ZnSe (zinc selenide). While the maximum temperature elevation of the detector chip is only less than 20 K as comparison.

The large differences in temperature elevation between the optical window and the detector chip are mainly caused by two reasons. Firstly, most of the energy of the laser with wavelength 808nm is absorbed by the window glass while a little proportion arrives at the detector chip. In other words, the heat flux produced by laser illumination in the chip is much smaller than that in the optical window. Secondly, under the action of strong heat conduction between the chip and the inner cylinder of the Dewar, the cooling rate of liquid nitrogen nearly balances with the heating rate of laser illumination for the detector chip. It is demonstrated in detail by computational results that when the laser power is 27 W, the value of peak heat flux on the chip's exterior surface along thickness at the moment t=2s is about one ninth of that on the exterior surface of the optical window, as shown in Fig. A4 (a) and Fig. A4 (b). And as indicated by Fig. A4 (c) and Fig. A4 (d), the peak heat flux on the chip's interior surface along thickness at the moment t=2s is only about 7% smaller than that on the exterior surface. i.e. most of the laser energy absorbed by the detector chip is diffused by heat conduction between the detector chip and the brass substrate.

According to the computational results at the end of laser illumination(t=5s), high stress appears in the components directly illuminated by laser like the optical window and the detector chip. Additionally, it also appears in regions of material discontinuity and geometrical discontinuity with high temperature gradient and mismatch in thermal expansion therein. As shown in Fig. 6(d) and Fig. A5, there are relatively high equivalent stress in regions including the detector chip, the optical window, the bottom and top ends of the Dewar's inner cylinder. According to Fig. 6(e) and Fig. 6 (f), the spatial peak equivalent stress of the optical window is situated at the contact interface between the window and its frame, and that of the detector chip is situated at contact interface between the chip and its substrate. Obviously, the numerical singularity influences the results greatly in regions with material discontinuity or geometrical discontinuity. Therefore, centroids of surfaces in the optical window and the detector chip are selected for analysis in present work to avoid the influence of stress singularity. As a complement, the spatial distributions of equivalent stress along the direction of thickness and width are also analyzed for these two components when the detector is illuminated by laser with different power from 1W to 27W.

As for the optical window, it is revealed that the temperature at the centers of both the exterior and interior surface always increases monotonically with time during illumination with laser power varying from 1W to 27W. As the laser power increases gradually, higher peak temperature would always be reached at the exterior surface center of the optical window. And it could be observed from Fig. 7(a) and Fig. 7(b) that the temperature changing rate at the exterior surface center is higher than that of the interior surface center during illumination. Therefore, the temperature difference between the exterior and interior surface center also increases with increasing the laser power.

It is further indicated that temperature varies at a highest rate within the time range from 0s to 1s. According to results for the case of the 27W laser power, the temperature changing rate of the exterior center is 1000K/s, about twice as large as that of the interior surface of 452K/s. After being illuminated for about 2s, the temperature changing rate of these two points both slows down obviously with continuing illumination. The temperature difference between these two points adds up to 472K at the moment of shutting-off laser (t=5s). In comparison with the optical window, the temperature difference between the exterior and interior surface centers in the detector chip is much smaller, which never exceeds 10K when the laser power varies from 1W to 27W.

In the cooling stage after illumination, the time-history of temperature of detector chip as shown in Fig. 7(c) and Fig. 7(d) differs obviously from that of the optical window. This difference might be resulted from the following facts. For the optical window, it relies on convective heat transfer on the external surface and heat conduction in heat dissipation, which makes it take at least 2~3s to cool down to the temperature level before illumination. For the detector chip, it is connected with the low-temperature environment by the material of high thermal conductivity(brass), so the temperature returns to the level in the pre-illumination state much more abruptly with the laser energy dissipated rapidly by the large heat sink.

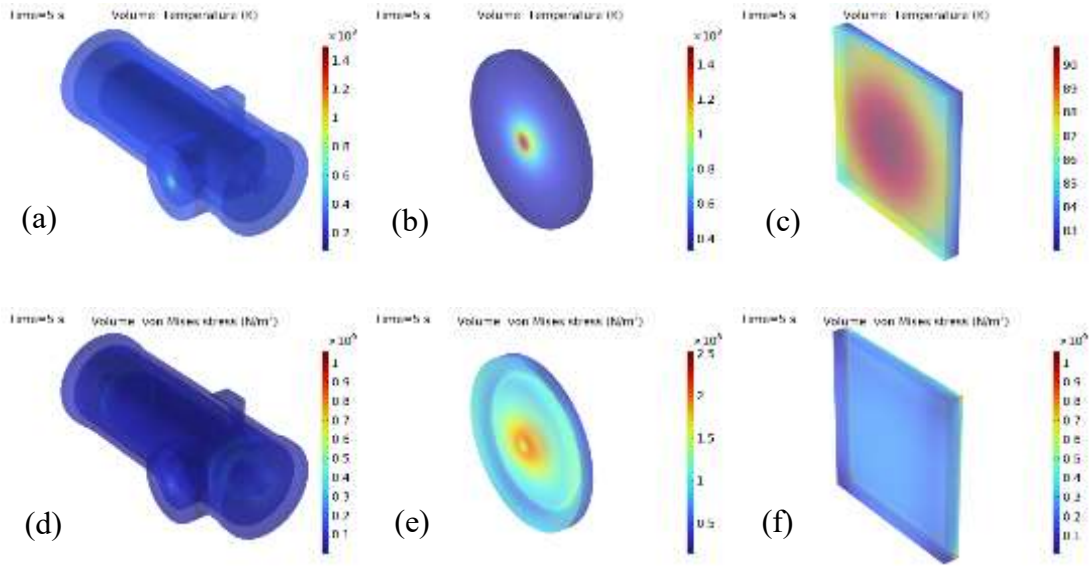

Fig. 6 Contour maps of temperature and stress under illumination by 27W laser (t=5s): (a) temperature of the whole structure, (b) temperature of the optical window, (c) temperature of the detector chip. (d) stress of the whole structure, (e) stress of the optical window and (f) stress of the detector chip.

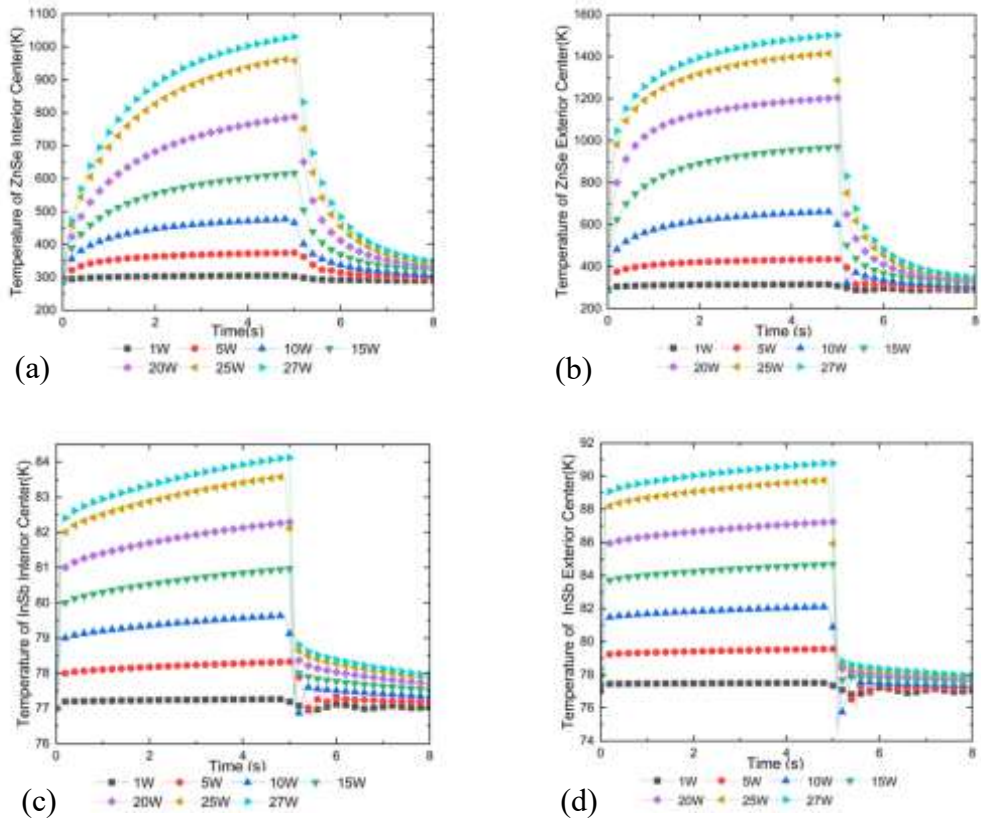

Fig. 7 Time dependent variation of the temperature of the of the optical window and the detector chip with the laser power ranging from 1W to 27W: (a) interior surface center of the optical window, (b) exterior surface center of the optical window, (c) interior surface center of the detector chip, (d) exterior surface center of the detector chip.

The temporal evolution characteristics of the volume averaged equivalent stress of the optical window shows a monotonic increasing trend during illumination(0~5s), as shown in Fig. A6. When the laser is turned off, the volume averaged equivalent stress starts to decrease. As a result, the peak value of the volume averaged equivalent stress appears at the moment of shutting-off laser(t=5s). The peak equivalent stress in the optical window grows gradually with the laser power increasing from 1W to 27W. Moreover, the peak equivalent stress during illumination with the laser power 27W is about ten times as large as that of the case with the laser power 1W. Such monotonic increasing of equivalent stress is also appropriate for most regions except for the central area on the exterior surface of the optical window when the detector is illuminated by laser with a relatively high power, as shown in Fig. 8(a) and Fig. 8(b). The equivalent stress of the centroid on the exterior surface achieves the peak value after being illuminated for about 1s when the laser power exceeds 25W. After that, it decreases gradually with successive illumination. When the adopted laser power is high enough, this phenomenon tends to exist in a very small range on the exterior surface of the optical window, whose radius is shorter than one tenth of the optical window's radius. Such unexpected decrease of equivalent stress in the optical window during the process of illumination under high power should be due to the fact that its temperature has reached an inflection point from which the thermal conductivity of zinc selenide (ZnSe) begins to increase as shown in Fig. A3. If the laser power density absorbed by the optical window is given, the temperature gradient would therefore reduce due to the increase of thermal conductivity. As a consequence, the equivalent stress in high-temperature regions of the optical window turns out to be decreasing with time during illumination.

On the contrary, the volume averaged equivalent stress in the detector chip decreases monotonically during illumination(0~5s), as shown in Fig. A7. It reaches the valley value at the moment of shutting-off laser(t=5s), and begins to rise at that moment, but it never exceeds the magnitude in the pre-illumination state. Such temporal evolution characteristic of the equivalent stress is almost the same for all the points on the detector chip. Although, the variation of the equivalent stress at the exterior surface center is more obvious than that of the interior surface, as shown in Fig. 8(c) and Fig. 8(d). The relative change of the equivalent stress during illumination is less than 5% of the value in the pre-illumination state when the laser power is 27W.

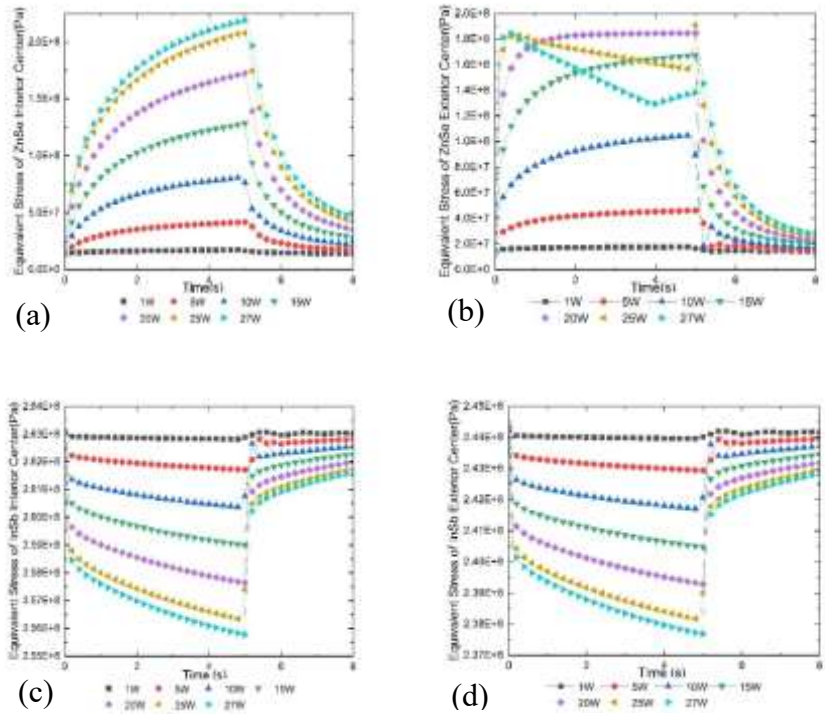

Fig. 8 Time dependent variation of the equivalent stress of the optical window and the detector chip with the laser power ranging from 1W to 27W: (a) interior surface center of the optical window, (b) exterior surface center of the optical window, (c) interior surface center of the detector chip, (d) exterior surface center of the detector chip.

As shown in Fig. 9(a) and Fig. 10(a), the path from the centroid of the exterior surface to that of the interior surface is selected for analyzing the variations along the thickness direction. For analysis along width direction, the path from centroid to the edge (or corner) is chosen for both exterior and interior surface.

According to the results of the optical window, the equivalent stress upon the interior surface increases at a faster rate than the exterior surface during illumination and therefore the comparative difference of the exterior surface and surface is reversed with the continuing of illumination. Before the instant t=1.8s, the equivalent stress on the exterior surface is higher than that on the interior surface. However, it turns out that the equivalent stress on the interior surface exceeds the exterior surface after 1.8s and late in the course of illumination, as shown in Fig. 9(b). The spatial distribution of the stress along width direction also varies with time. As shown in Fig. 9(c) and Fig. 9(d), the equivalent stress varies a little along width direction both on the exterior and interior surface in the pre-illumination and post-illumination state (5-8s). During illumination, the equivalent stress level increases as a whole with time. And it grows faster in certain regions near the centroid and the edge, resulting in two peaks in corresponding regions. But the comparative difference of the two peaks are distinctive on the exterior surface and the interior surface. On the exterior surface, the peak equivalent stress near the centroid is two times as large as the one near the edge at the shut-off time of the laser illumination (t=5s). While, the magnitudes of the two peaks are closer to each other at the same time (t=5s) on the interior surface.

The distribution of the equivalent stress in the detector chip barely changes in the time range from 0s to 8s along both width and thickness direction. The equivalent stress continually increases from the exterior surface center to the interior surface center, as shown in Fig. 10(b). Along the direction of width, it decreases continuously from the centroid to the corner on the exterior surface but increases continuously from the centroid to the corner on the interior surface, as indicated by Fig. 10(c) and Fig. 10(d). It could be deduced that the characteristic of spatial distribution of the equivalent stress in the detector chip changes slightly with time during illumination by laser with power 27W.

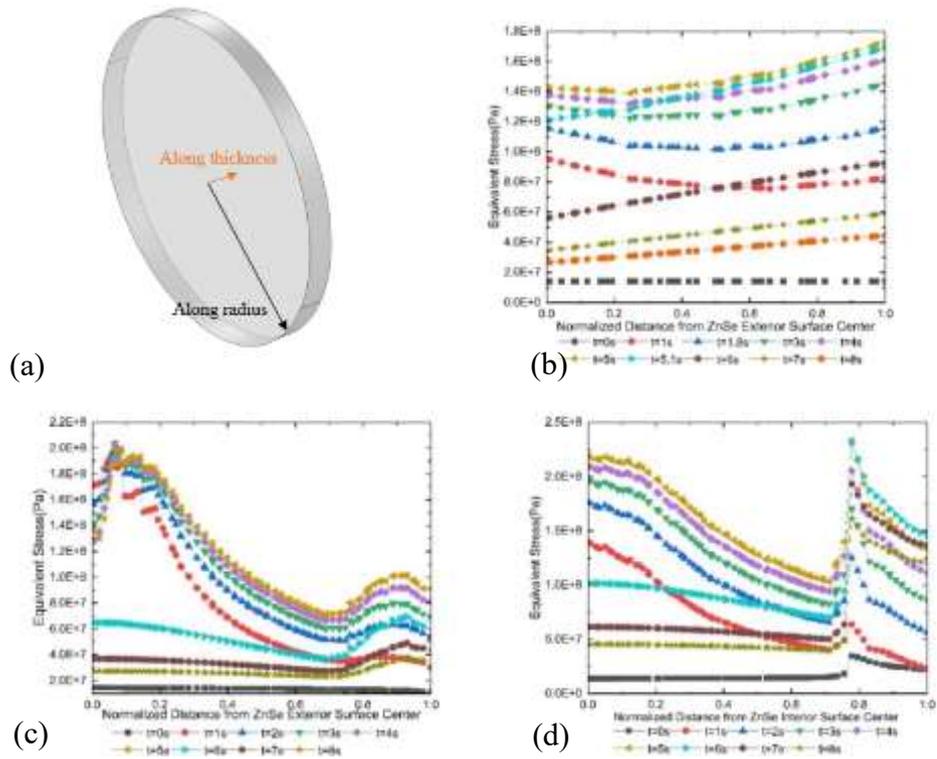

Fig. 9 Equivalent stress distributions of the optical window projected along different directions: (a) schematic diagram of projection directions, (b) projection along thickness from the exterior surface center to the interior surface center, (c) projection along the radial direction from the exterior surface center to the exterior surface edge (d) projection along the radial direction from the interior surface center to the interior surface edge.

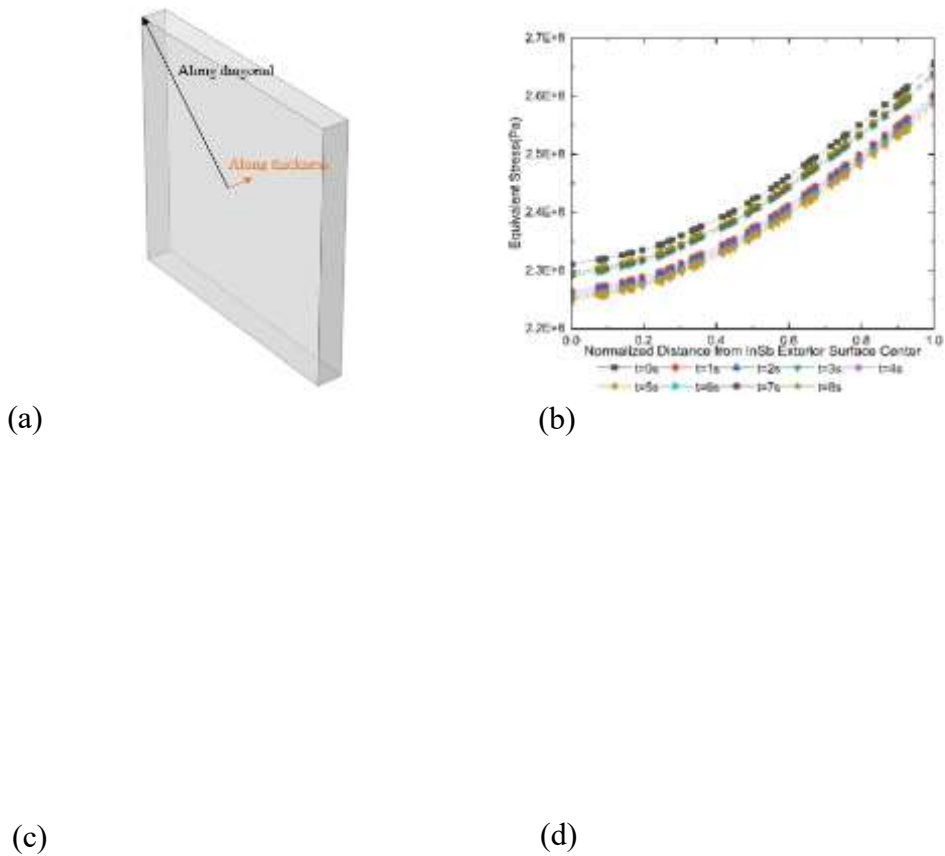

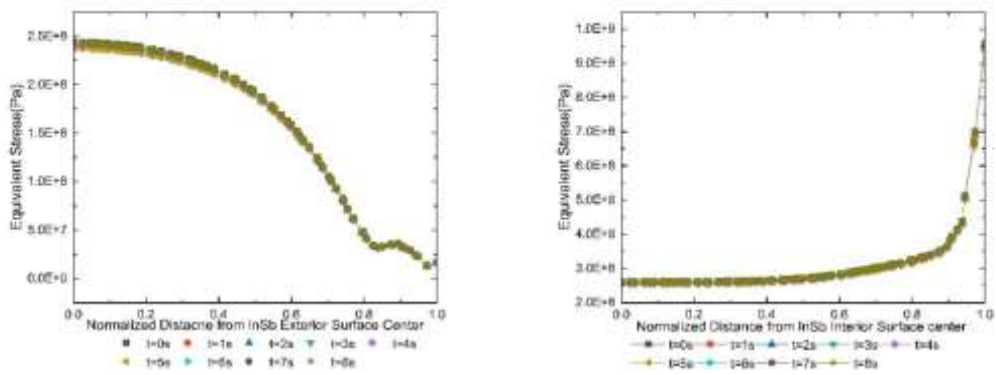

Fig. 10 Equivalent stress distributions of the detector chip projected along different directions: (a) schematic diagram of projection directions, (b) projection along the thickness direction from the exterior surface center to the interior surface center, (c) projection along the diagonal direction from the exterior surface center to the exterior surface corner, (d) projection along the diagonal direction from the interior surface center to the interior surface corner.

When the laser power equals to 10W, the dependence of thermomechanical outcomes on the velocity of forced convection outside the optical window is also discussed in present work. The velocity of forced convection ranges from 0.1 m/s to 100 m/s. It is indicated that the optical window is the mostly affected component by the variation of convection velocity. The peak temperature of the centroid both on the exterior and interior surface of the optical window during illumination (0~5s) decreases when the velocity varies from 0.1 m/s to 100 m/s, as shown in Fig. 11 (a) and Fig. 11 (b). At the same time, the peak equivalent stress of the surface center on the optical window during illumination (0~5s) also decreases with the increasing velocity of forced convection. Moreover, the equivalent stress of the interior surface center reduced more apparently than that of the exterior surface center, as shown in Fig. 11(c) and Fig. 11(d). The peak value of equivalent stress of the exterior surface center in the case of 100 m/s is reduced by around 10% compared to that for the case of 0.1m/s.

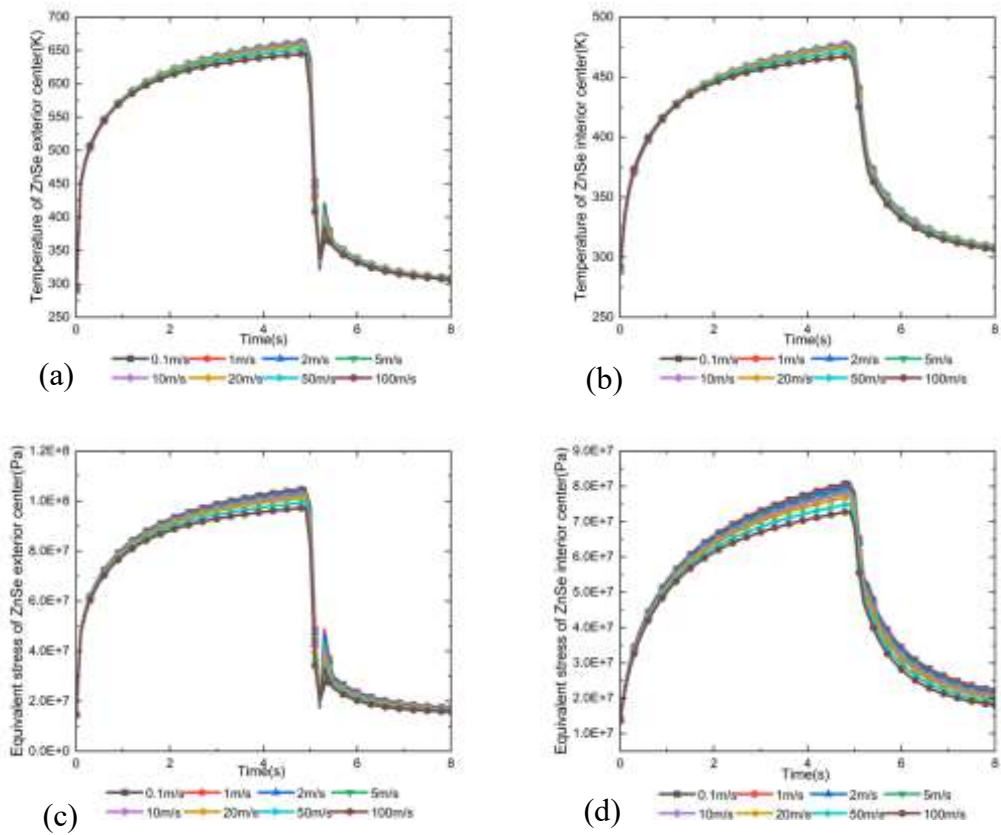

Fig. 11 Time-dependent temperature and equivalent stress curves of the optical window under illumination of 10W laser with velocity of forced convection ranging from 0.1m/s to 100 m/s: (a) temperature of the exterior surface center, (b) temperature of the interior surface center, (c) equivalent stress of the exterior surface center (d) equivalent stress of the exterior surface center.

## 4.Conclusions

The primary damage mechanism of a cooled infrared detector under laser illumination has been identified through a comprehensive analysis of its thermomechanical behavior, combining experimental investigations with numerical simulations. The experimental results demonstrate that the failure of the cooled infrared detector under laser illumination is primarily driven by optical window damage, manifested in two key aspects: (1) fracture of the window glass induced by excessive thermal stress, which results in the collapse of the vacuum layer, and (2) severe degradation of the optical view due to reduced transmittance caused by ablation of the films on the window glass's exterior surface. Differing from the outcomes documented in earlier studies, there is no significant damage observed in the detector chip under optical microscopy or scanning electron microscopy. In other words, the laser absorptivity of the optical window is significantly enhanced by pollutants on its exterior surface, leaving a smaller proportion of the laser being absorbed by the detector chip. The resulting high temperature elevation during laser illumination can cause the window material to melt and fracture, leading to the collapse of the vacuum layer. This prevents the target signals from reaching the detector chip, ultimately resulting in permanent failure to the detector.

Unlike previous studies that predominantly focused on the failure modes of the detector chip or the optical window alone, the current numerical model integrates the entire detector system, including the Dewar, chip, and optical window, as a unified entity. This model encompasses both the cooling process and the failure progression of the detector from a time-scale perspective. The computational results reveal that the temperatures of both the optical window and the detector chip exhibit a rapid increase under laser illumination. The equivalent stress in the optical window generally rises, whereas it decreases in the detector chip. This contrasting behavior can be attributed to the significant differences in their initial and boundary conditions. Before illumination, the detector chip is subjected to high tensile stress due to the liquid nitrogen cooling process. During illumination, thermal expansion of the chip partially offsets the initial cooling effects, causing a gradual decline in equivalent stress with prolonged illumination duration.

The temporal evolution and spatial distribution of the detector's thermomechanical responses to laser illumination are significantly influenced by the power density. Due to the nonlinear relationship between thermophysical parameters and temperature, the increasing rate of equivalent stress at the center of the exterior surface gradually slows down as the laser power rises. With the laser power density ranging from $3.18 \times$

$10^5 W/m^2$ to $8.59 \times 10^6 W/m^2$, The peak temperature elevation of the exterior surface center in the optical window increases by over 40 times, and the peak equivalent stress of the interior surface center in the optical window increases by 9 times. With the velocity of forced convection ranging from 0.1 m/s to 100 m/s, the maximum temperature elevation decreases by about 6%, the peak equivalent stress of the interior surface center of the optical window decreases by about 10%.

In summary, the failure mechanism of the detector under laser illumination has been elucidated in this study, primarily attributed to the collapse of the vacuum layer resulted from cracking of the optical window. These findings provide a solid foundation for optimizing the thermomechanical design to enhance the detector's survivability under extreme conditions.

## Acknowledgement

This work was supported by the National Natural Science Foundation of China (Grant No. 11572327 and 12232020).

Appendix

Figure A1 shows a photograph of the window glass that has not undergone the laser irradiation test.

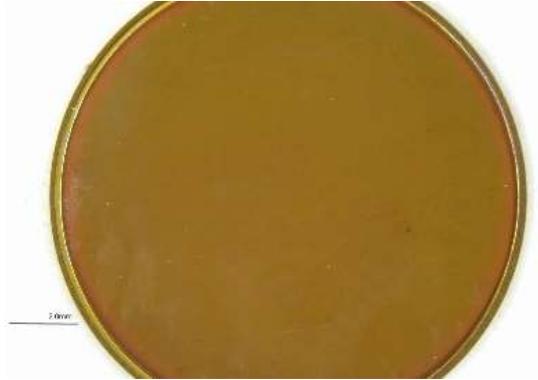

Fig. A1 The optical window before damage

Figure A2 is the photograph of the detector that display partially its inner detail with the removal of the optical window.

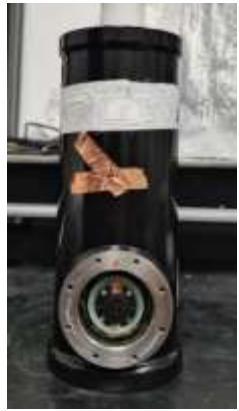

Fig. A2 The realistic structure of the detector

Figure A3 shows the temperature dependences of the parameters including density, heat capacity, thermal conductivity, thermal expansion coefficient, elastic modulus and Poisson's ratio.

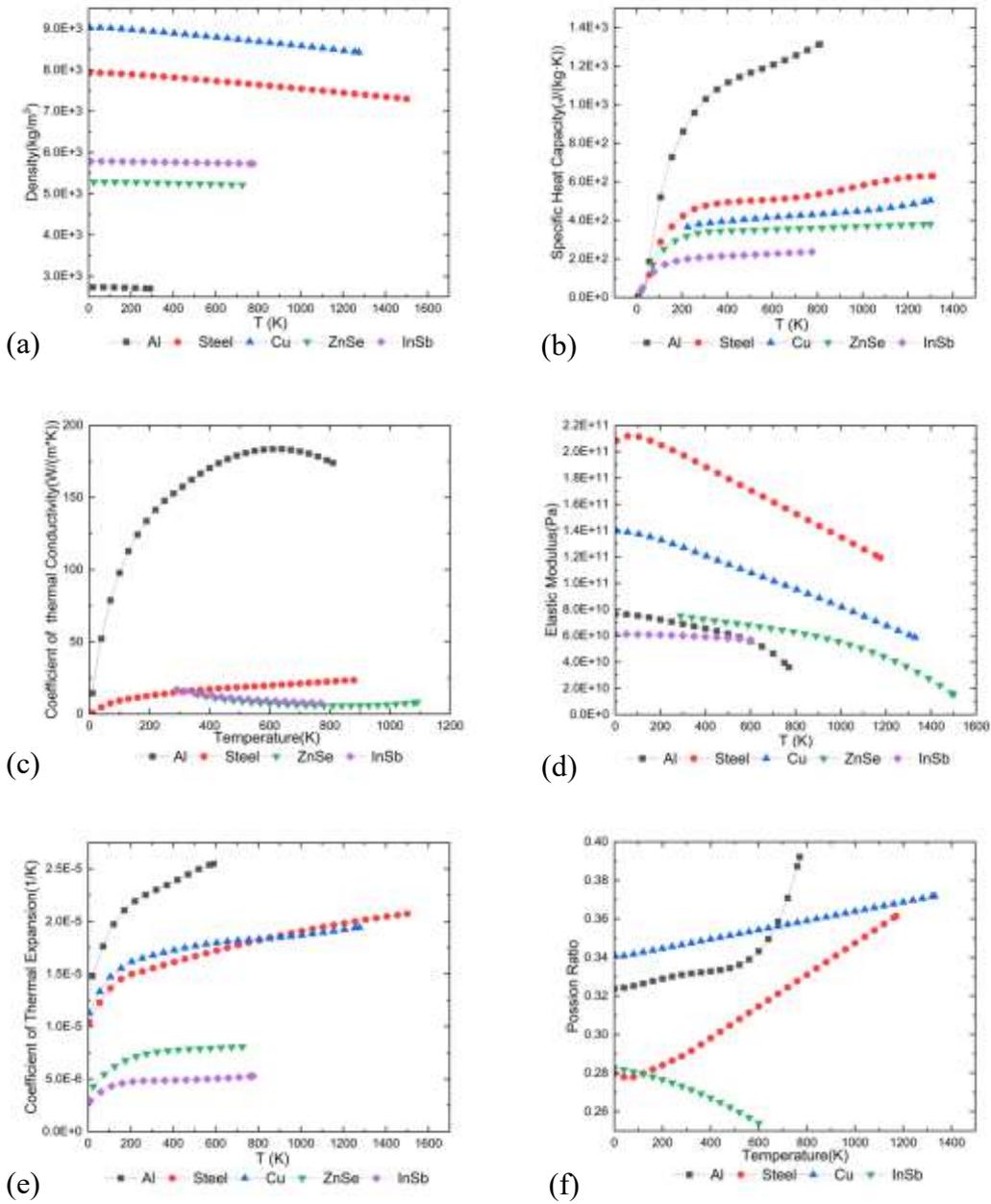

Fig. A3 Temperature-dependent parameters variation in different materials applied.

(a)Density (b) Heat capacity (c) Coefficient of thermal conductivity (d) Elastic Modulus (e) Coefficient of thermal conductivity (f) Possion's ratio [55]

The governing equation [57] of heat conduction which determines the temperature field of the detector structure could be derived based on Fourier's law of heat conduction and the first law of thermodynamics as

$$\rho C \frac{\partial T}{\partial t} = \nabla \cdot (k \nabla T) \tag{Eq.1}$$

Wherein, the expression on the left-hand side equals to 0 in the first computational step when solve the temperature field for the steady thermodynamic state of the detector cooled with liquid nitrogen. That is to say, the distribution of temperature field does not vary with time, so the equation could be degenerated into Laplace's equation.

In the numerical modeling, the energy transfers mainly through the boundary surfaces of the detector, including the internal surfaces of the inner cylinder which are directly in contact with the liquid nitrogen, the exterior surfaces of the optical window and detector chip which are directly illuminated by the laser, the exterior surfaces of the outer cylinder on which the detector exchange heat with surrounding air, etc. Heat flux will be exerted on the boundary surfaces mentioned above corresponding to the thermodynamic boundary conditions similar to the in-situ test. It should be noted that the intensity of radiative heat transfers between the outer cylinder and inner cylinder on each sides of the vacuum layer is small enough to be ignored in the model.

In detail, constant temperature boundary is applied on the interior surfaces of the inner cylinder as it keeps direct contact with the liquid nitrogen for the sake of simplification.

$$T = T_0 \tag{Eq.2}$$

The energy consumed on producing photoelectric effect in form of the increasing open-circuit voltage takes up only a small proportion of laser energy illuminated on the detector. Therefore, laser power absorbed by the detector is approximately treated as an equivalent Gaussian heat flux, which is proportioned before being applied to the exterior surfaces of the optical window, the detector chip and the substrate of the chip. The equivalent heat flux is proportioned according to the absorption rate and transmittance of the optical window, together with the absorption of the detector chip and substrate. In this numerical model, the absorption rate and transmittance of the optical window are calculated according to the multilayer optical medium theory and experimental measurements respectively. The thermal boundary conditions on the laser illuminated surface can be expressed as follows,

$$-\mathbf{n} \cdot \mathbf{q} = q_0 \quad \text{(Eq.3)}$$

The convective heat transfer boundary conditions between surrounding air and the outmost top and side surfaces of the Dewar, and the exterior surface of the optical window are,

$$-\mathbf{n} \cdot \mathbf{q_1} = q_1 \quad \text{(Eq.4)}$$

$$q_1 = h(T_{ext} - T) \quad \text{(Eq.5)}$$

where the convective heat transfer coefficient h depends on the flowing velocity of the air. The radiative heat transfer boundary condition on the outmost surfaces of the detector is,

$$-\mathbf{n} \cdot \mathbf{q_2} = \varepsilon\sigma(T_{amb}^4 - T^4) \quad \text{(Eq.6)}$$

The governing equations [58] which control the deformation fields of the whole structure in the detector could be derived by combining the basic equations of elasticity with considering the contribution of thermal expansion as,

$$\frac{3(1-v)}{1+v}\nabla(\nabla \cdot \boldsymbol{u}) - \frac{3(1-2v)}{2(1+v)}\nabla \times (\nabla \times \boldsymbol{u}) = \alpha(\nabla T) \quad \text{(Eq.7)}$$

The relationships between the stress and deformation for every component in the detector are,

$$\sigma_{ij} = C_{ijkl}(\frac{1}{2}(u_{k,l} + u_{l,k}) - \alpha\Delta T \delta_{ij}) \quad \text{(Eq.8)}$$

With regard to the mechanical boundary conditions, the rigid body displacement of the whole structure is set to 0. And it is assumed that there is no deformation nor stress in the whole structure at room temperature in the initial state.

Table A1 list the nomenclatures used in the equations (Eq.1) ~ (Eq.8).

Table A1 Nomenclatures

| | |
|---|---|
| $\rho$ | Density |
| C | Heat capacity |
| T | Temperature |
| k | Thermal conductivity |
| $T_0$ | Temperature of the liquid nitrogen |

| **n** | Unit normal vector |
|---|---|
| $q_0$ | Heat flux equivalent to laser |
| $q_1$ | Heat flux of natural convection |
| $h$ | Convective heat transfer coefficient |
| $T_{ext}, T_{amb}$ | External environment temperature, ambient temperature |
| $\varepsilon$ | Surface emissity |
| $\sigma$ | Stefan-Boltzmann constant |
| $v$ | Possion |
| $\sigma_1, \sigma_2, \sigma_3$ | The first, second and third principal stress |
| $\sigma_e$ | $= \sqrt{\frac{1}{2}((\sigma_1 - \sigma_2)^2 + (\sigma_2 - \sigma_3)^2 + (\sigma_1 - \sigma_3)^2)}$, Von-Mises equivalent stress |
| $\sigma_{ij}$ | Stress tensor |
| $C_{ijkl}$ | Elasticity tensor |
| $u_{k,l}, u_{l,k}$ | Strain tensor |
| $\delta_{ij}$ | Kronecker symbol, which equals to 1 when $i = j$ and 0 when $i \neq j$ |

Figure A4 depicts the contour maps of the x-component of total heat flux at the instant t=2s with x being the direction along thickness of both the optical window glass and the chip.

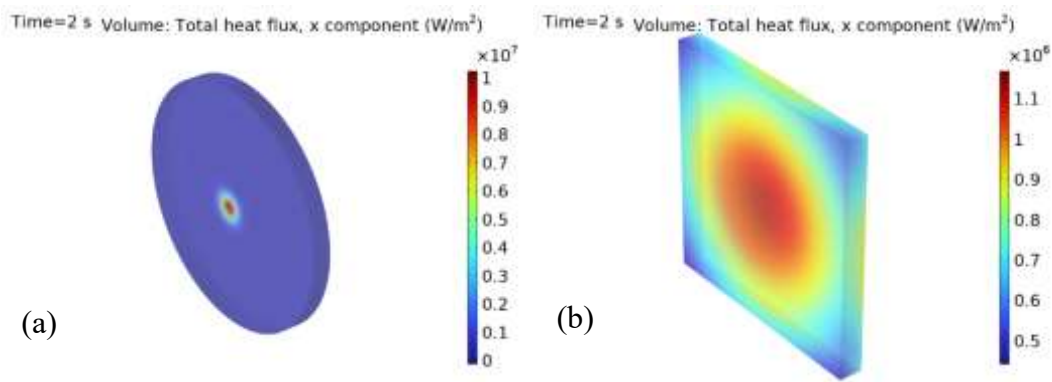

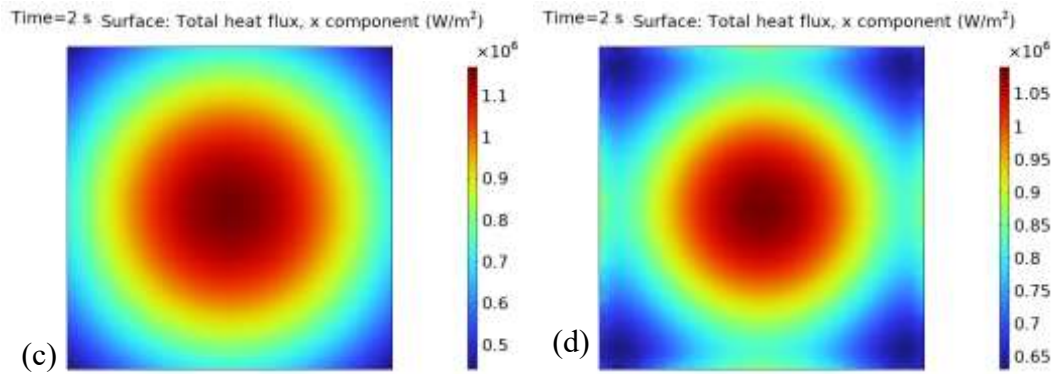

(c)                                                                (d)

Fig. A4 contour maps of x component of total heat flux (along thickness) of the optical window and the detector chip under illumination by a 27W laser (t=2s)

(a) The optical window (b) The detector chip (c) The exterior surface of the chip (d) The interior surface of the chip

Figure A5 shows the contour map of equivalent stress in the inner cylinder of the Dewar at the instant of t=5s when the illumination is shut off.

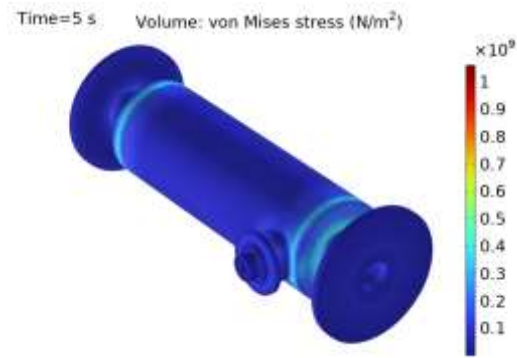

Fig. A5 contour map of equivalent stress in the inner cylinder of the Dewar at the end of illumination (t=5s)

Figure A6 and A7 show the time history of the volume averaged equivalent stress of the optical window and the detector chip with the laser power ranging from 1W to 27W, respectively.

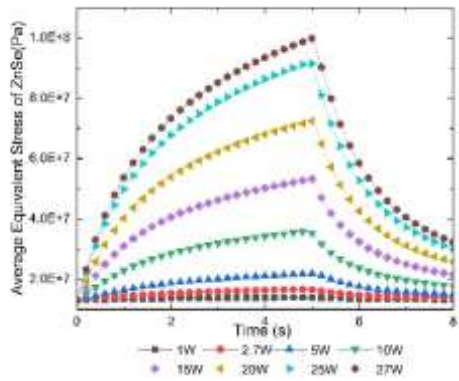

Fig. A6 Time-dependent variation of volume averaged equivalent stress of the optical window with the laser power ranging from 1W to 27W

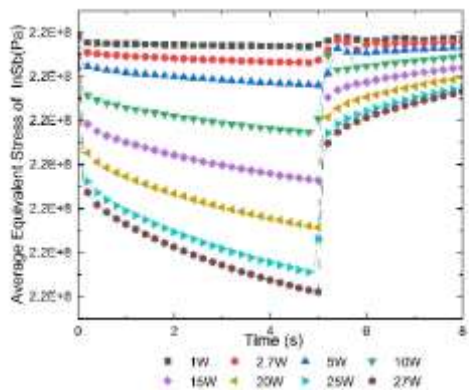

Fig. A7 Time-dependent variation of volume averaged equivalent stress of the detector chip with the laser power ranging from 1W to 27W